\documentclass{aa}
\usepackage{graphicx}
\usepackage{txfonts}
\newcommand{\NaI}{\ion{Na}{I}}
\newcommand{\CaI}{\ion{Ca}{I}}
\begin{document}
\title{Optical spectropolarimetry of V4332 Sagittarii\thanks{Based on observations made with 
ESO Telescopes at the Paranal Observatory under programme ID 087.D-0019(A). 
Processed data are available in electronic form at the CDS via anonymous ftp to 
cdsarc.u-strasbg.fr (130.79.128.5) or via http://cdsweb.u-strasbg.fr/cgi-bin/qcat?J/A+A/.}
}
\author{
T. Kami\'nski\inst{1}
\and 
R. Tylenda\inst{2} 
}  

%\offprints{R. Tylenda}
\institute{
            Max-Planck Institut f\"ur Radioastronomie, Auf dem
            H\"ugel 69, 53121 Bonn, Germany\\
            \email{kaminski@mpifr.de}
       \and 
           Department for Astrophysics, N. Copernicus
            Astronomical Centre, Rabia\'{n}ska 8,
            87-100 Toru\'{n}, Poland\\ 
            \email{tylenda@ncac.torun.pl}}
\date{Received; accepted}
\abstract
{The eruption of V4332 Sgr was observed in 1994. During the
outburst, the object became extremely red, so it has been considered as
to belong to red transients of the V838~Mon type. Optical spectroscopy
obtained a few years after the eruption showed a faint M-type stellar
spectrum underlying numerous molecular and atomic emission features.
It has recently been suggested that the central object in V4332~Sgr is now
 hidden in a dusty disc and that the photospheric spectrum of this object
observed in the optical results from scattering of the radiation of 
the central star on dust grains in the circumstellar matter. 
Recent polarimetric photometry has shown that the
optical radiation of the object is indeed strongly polarized.}
{If it results from scattering on
dust grains, the optical continuum of the object
is expected to be polarized. The emission features, on the contrary --
as observed directly from circumstellar regions -- are expected to be unpolarized. 
We investigate these predictions.} 
{We present and analyse our spectropolarimetric observations of V4332~Sgr
obtained with the VLT in the optical region.}
{The optical continuum of V4332~Sgr is linearly polarized with a typical degree of
16.6\%. A clear depolarization is observed in the spectral regions where emission 
features contribute significantly to the observed flux.
The only prominent exception is the \CaI\,$\lambda$6573 emission line, which is
polarized in 21\%.}
{The results of our spectropolarimetric observations are in accord with the 
proposed geometry of the system and the advocated nature of the observed optical spectrum of 
V4332~Sgr. The continuum is not seen directly but results from scattering on dust 
within the disc and polar outflow, while most of the gas emission comes from the polar outflow 
excited by the radiation field of the central source. 
Additionally, the observed polarization patterns suggest a stratification of the outflow.
}
        
\titlerunning{Polarization of V4332 Sgr}
\authorrunning{Kami\'nski \& Tylenda}
\maketitle
%----------------------------------------------------------------------------------
\section{Introduction}
 The eruption of V4332~Sgr was observed in 1994 \citep{martini}.
Discovered as a possible classical nova, the object appeared to be unusual,
mainly because of its spectral evolution. 
It quickly evolved from K to M spectral types and after a month declined 
to a very late M-type giant. The object is now classified as a red 
transient of the V838~Mon type \citep{muna,tcgs}.
As discussed by \citet{ts06},
thermonuclear mechanisms (classical nova, He-shell flash)
cannot explain the
observed outbursts of these objects. The stellar
collision-merger scenario proposed in \citet{st03} and
further developed in \citet{ts06} is the most promising
hypothesis for explaining the nature of these eruptions. 
Strong arguments supporting this idea came from \citet{thk11}, who
showed that the eruption of V1309~Sco, another red transient, resulted from
a merger of a contact binary.
 
More than a decade after outburst, V4332~Sgr continues
to present unusual observational characteristics. Apart from displaying an M-type
continuum in the optical,
it shows a unique emission-line spectrum with very low
excitation \citep{tcgs,kimes}.
The object is also very bright in the infrared \citep{tcgs,baner07}.

After having analysed the optical emission-line
spectrum and the spectral energy distribution of V4332~Sgr, \citet{kst}
conclude that the main object, probably an M-type giant, is hidden in a
circumstellar disc seen almost edge-on. The stellar-like spectrum observed
in the optical probably results from scattering of the light of the central star on
dust grains at the outer edge of the disc. This implies that the optical
spectrum should display significant polarization. Subsequent
polarimetric measurements done by \citet{kt11} in the $V$ and $R$
photometric bands have shown that the optical light from V4332~Sgr is
indeed strongly polarized (26\% and 11\%, respectively). 
As proposed in \citet{kst}, the
emission-line spectrum results from radiative excitation of
atoms and molecules
by strong radiation of the (hidden for us) central object. Thus,
unlike the observed continuum, the emission lines should be
unpolarized. To verify this hypothesis, we have obtained spectropolarimetric
measurements of V4332~Sgr in the optical. The results and their analysis are
presented in this paper.

%++++++++++++++++++++++++++++++++++++++++++++++++++++++++++++++++++++
\section{Observations and data processing  \label{obs_sect}}

V4332~Sgr was observed with the Focal Reducer and Low Dispersion Spectrograph 2 (FORS2) 
attached to the Very Large Telescope. The observations were performed on 2011 May 18 in 
the linear-polarization mode in which a retarder plate is rotated in front of 
a Wollaston prism. Our source was observed in two cycles, each with four retarder 
angles (0\degr, 22\fdg5, 45\degr, and 67\fdg5) and an exposition time of 
290\,s per angle. We used the 1200R+93 grism, which with a slit width of 1\arcsec,
provided a spectral resolution of about $\lambda/\Delta \lambda$=2100 and 
a spectral coverage of 5800--7300\,\AA. Seeing during the observations was
1\farcs1--1\farcs2. The degree of instrumental polarization was controlled by 
observations of a zero-polarization standard star, WD\,2039-202. 

The data reduction was performed using the standard FORS2 pipeline. It included 
corrections for bias and flatfield, wavelength calibration, and extraction of 
spectra. The reduced Stokes parameters were computed from the signal extracted 
from spectra of the ordinary ($f^o$) and extraordinary ($f^e$) rays
\begin{equation}
P_Q=\frac{f(0\degr)-f(45\degr)}{2},\;\; 
P_U=\frac{f(22\fdg5)-f(67\fdg5)}{2},
\end{equation}
where
\begin{equation}
f(\theta_i)=\frac{f^o(\theta_i)-f^e(\theta_i)}{f^o(\theta_i)+f^e(\theta_i)}\,\;{\rm for~}\,\theta_i=0\degr, 22\fdg5, 45\degr, 67\fdg5.
 \end{equation} 
The total intensity (the Stokes $I$ parameter) was computed as the average of all 
the observed spectra. The degree of linear polarization $P_L$ and the polarization 
angle $P_A$ were computed as
\begin{equation}
P_L=\sqrt{P_Q^2+P_U^2},\\
P_A=\frac{1}{2}\arctan \biggl(\frac{P_U}{P_Q}\biggr)+\Theta_0,
\end{equation}
where
\begin{equation}
\Theta_0=\left\{ \begin{array}{ll ll l}
0        & \textrm{if} & P_Q >0&\textrm{and}& P_U \geq 0\\
180\degr & \textrm{if} & P_Q >0&\textrm{and}& P_U \leq 0\\
90\degr  & \textrm{if} & P_Q >0&.& 
\end{array} \right.
\end{equation}
The uncertainties in $P_L$ and $P_A$ were computed from $1\sigma$ statistical 
errors in the extracted spectra of the ordinary and extraordinary rays and 
propagated to $P_L$ and $P_A$ using formulae from \cite{fossati}. The results 
from the two cycles were averaged. The flux of V4332~Sgr turned out to be 
very weak, particularly at 5800--6000\,\AA\ and 6150--6350\,\AA. To improve 
the presentation of the data here, the spectra were significantly smoothed 
and the polarization parameters recalculated in the degraded resolution. 
Typically, we applied boxcar averaging within 13 spectral bins.

The derived polarization degree of the zero-polarization standard of 
$P_L$=(0.42$\pm$0.23)\% (weighted mean and 1\,rms of the residuals) agrees 
with the $R$-band catalogue value for this star \citep{FORSpipelineMan}, 
$P_L$=(0.24$\pm$0.45)\%, proving that the instrumental bias on $P_L$ is marginal. 
However, the polarization angle is known to be affected by instrumental 
effects \citep{FORSMan}, and the results were corrected for these effects 
using chromatic instrumental-zero points provided by the observatory 
\citep{FORSpipelineMan}. The amplitude of the correction varied between 
0 and 3\degr\ within the observed spectral range. Although no standard star 
with high polarization was observed during our observing run, constant 
monitoring of the instrumental polarization performed by 
the observatory\footnote{\url{http://www.eso.org/observing/dfo/quality/FORS2/reports/HEALTH/trend_report_PMOS_angle_HC.html}} 
assures that the applied correction is valid.

\section{Results  \label{res_sect}}

\begin{figure*}
\includegraphics[angle=270,width=\textwidth]{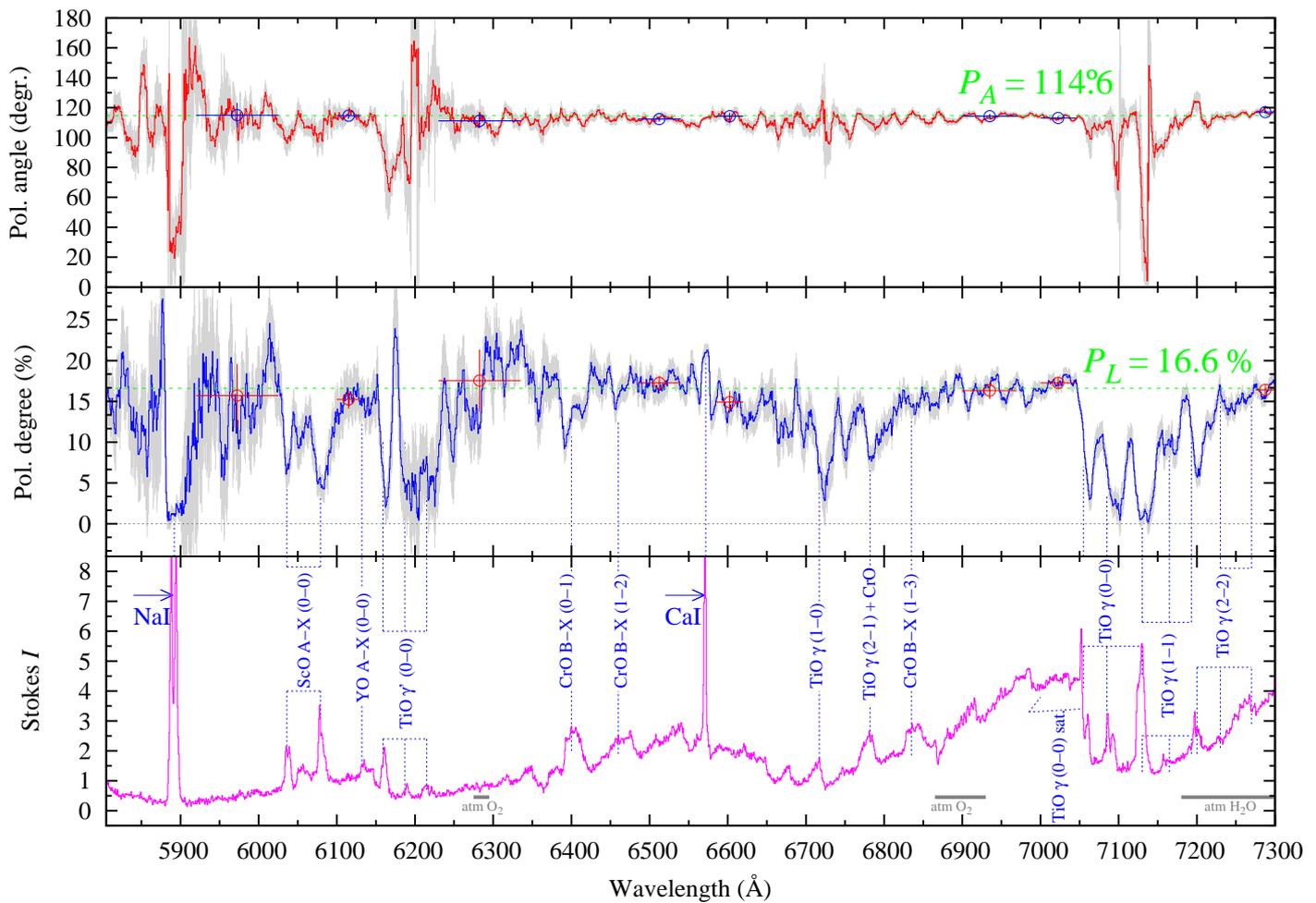}
\caption{Spectropolarimetry of V4332 Sgr. 
{\it Bottom panel}: the observed spectrum 
of the object (in arbitrary units) with the identification of the strongest emission features.
{\it Middle panel}: the degree of the observed linear polarization.
{\it Top panel}: the polarization angle. 
$1\sigma$ errors of the polarization angle
and degree are shown in grey in the top and bottom panels. 
Horizontal dashed lines in the middle and top
panels indicate mean polarization degree (16.6\%) and angle (114\fdg6) 
of the continuum as measured in the spectral ranges marked by the horizontal bars attached 
to the open circles. The circles denote average values of $P_A$ (top panel) or $P_L$ 
(middle panel) within the individual range and the vertical errorbars represent 
the corresponding one standard deviation.
%(see Fig.~\ref{cont_pol_fig}).
}\label{specpol_fig}
\end{figure*}

The results of our spectropolarimetric observations are shown in
Fig.~\ref{specpol_fig}. The spectrum of V4332~Sgr, displayed in the bottom
panel of the figure, 
covers several emission bands of metal oxides (TiO, CrO, ScO, and YO) 
and two features of atomic resonance lines, i.e.,
the marginally resolved doublet of \NaI\ near 5993\,\AA\   
and the intercombination line of \CaI\ at 6573\,\AA.
For a more detailed description and analysis of the
optical spectrum of V4332~Sgr, see \citet{kst}.
The degree of linear polarization as a function of 
wavelength is presented in the middle panel, while the top panel shows the
polarization angle. Flux-weighted means of the polarization degree and angle
calculated for the whole observed spectrum
are 15.3\%$\pm$6.6\% and 109\degr$\pm$25\degr, respectively. This can be
compared to 11.3\%$\pm$2.0\% and 116\degr$\pm$5\degr\ obtained in \citet{kt11}
from the photopolarimetry of V4332~Sgr in the $R$ band. (The
spectral region of the present observations roughly corresponds to that
covered by the $R$ band.)

\begin{figure*}
\includegraphics[angle=270,width=\textwidth]{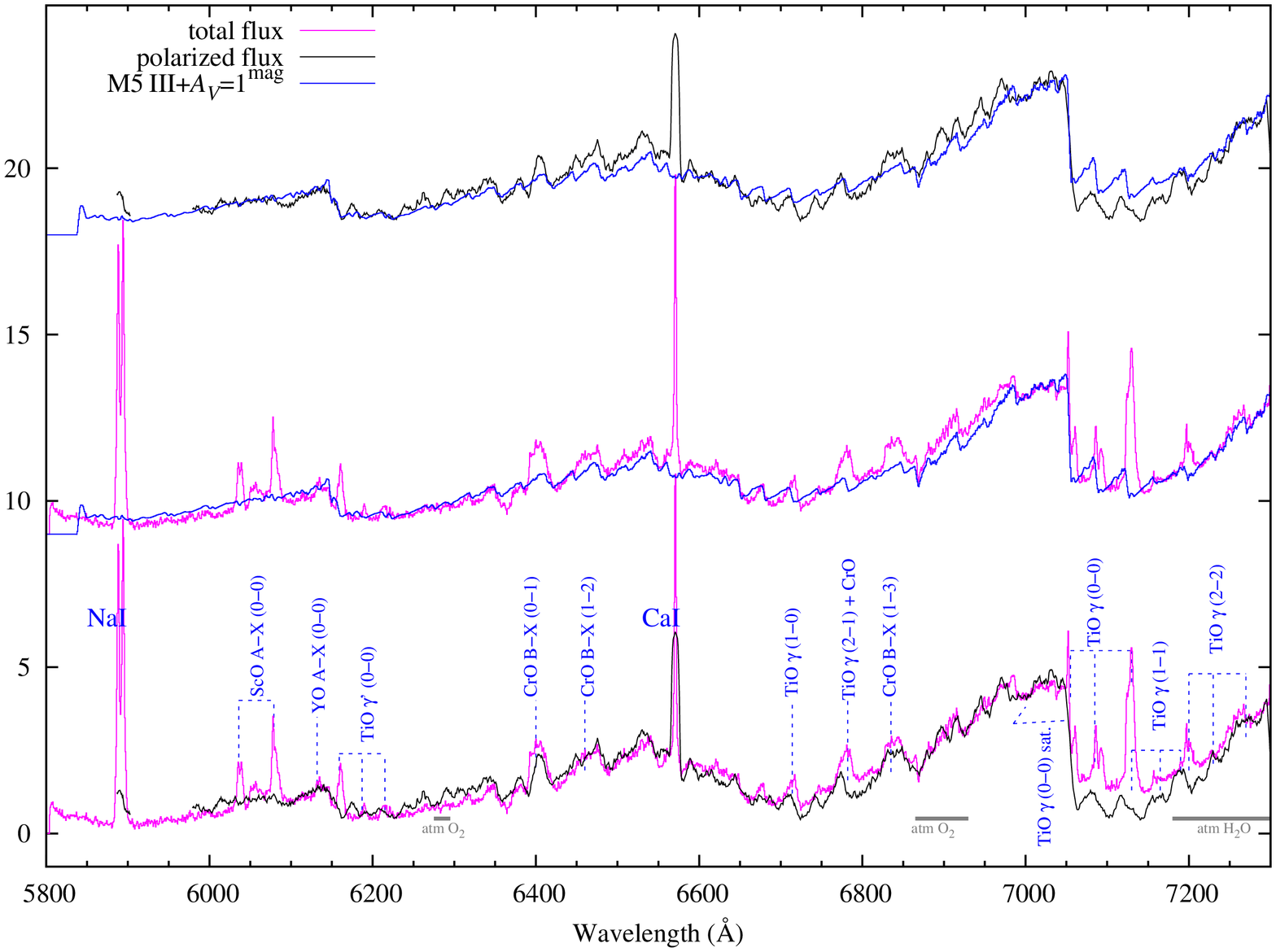}
\caption{Total and polarized spectral flux of V4332 Sgr plotted as the
magenta and black curves, respectively. A
standard M5\,III spectrum, reddened with $E_{B-V} = 0.32$, is also displayed
with the blue curve for a comparison.
}
\label{specpol2_fig}
\end{figure*}

As can be seen from the middle panel of
Fig.~\ref{specpol_fig}, the observed spectrum is strongly polarized.
A maximum polarization degree reaches $\sim$20\%. In certain
spectral regions, however, the polarization drops to almost zero
(the interstellar polarization towards V4332~Sgr is discussed in Appendix~\ref{ISP}).
When comparing the middle and bottom panels of Fig.~\ref{specpol_fig}, it is
evident that the depolarization takes place in spectral regions affected
or dominated by emission features. To better illustrate this, we 
calculated ``polarized flux'' defined as a product of the
polarization degree and the total observed flux. The result is displayed
in Fig.~\ref{specpol2_fig}, as
is a standard M5\,III spectrum reddened as V4332~Sgr, i.e., with 
$E_{B-V} = 0.32$ \citep{kst}. As can be seen from the upper part of
Fig.~\ref{specpol2_fig}, the spectrum of the polarized flux
shows an M5-type spectrum dominated by strong    
absorption bands of TiO.
The only prominent emission feature clearly seen in the polarized flux 
is the $\lambda$6573 line of \CaI, but the strongest bands of CrO are also present.
All other features seen in emission in the total-flux spectrum, 
in particular the \NaI\,D lines, are absent in the polarized spectrum.

\begin{figure*}
\includegraphics[angle=0,width=0.5\textwidth]{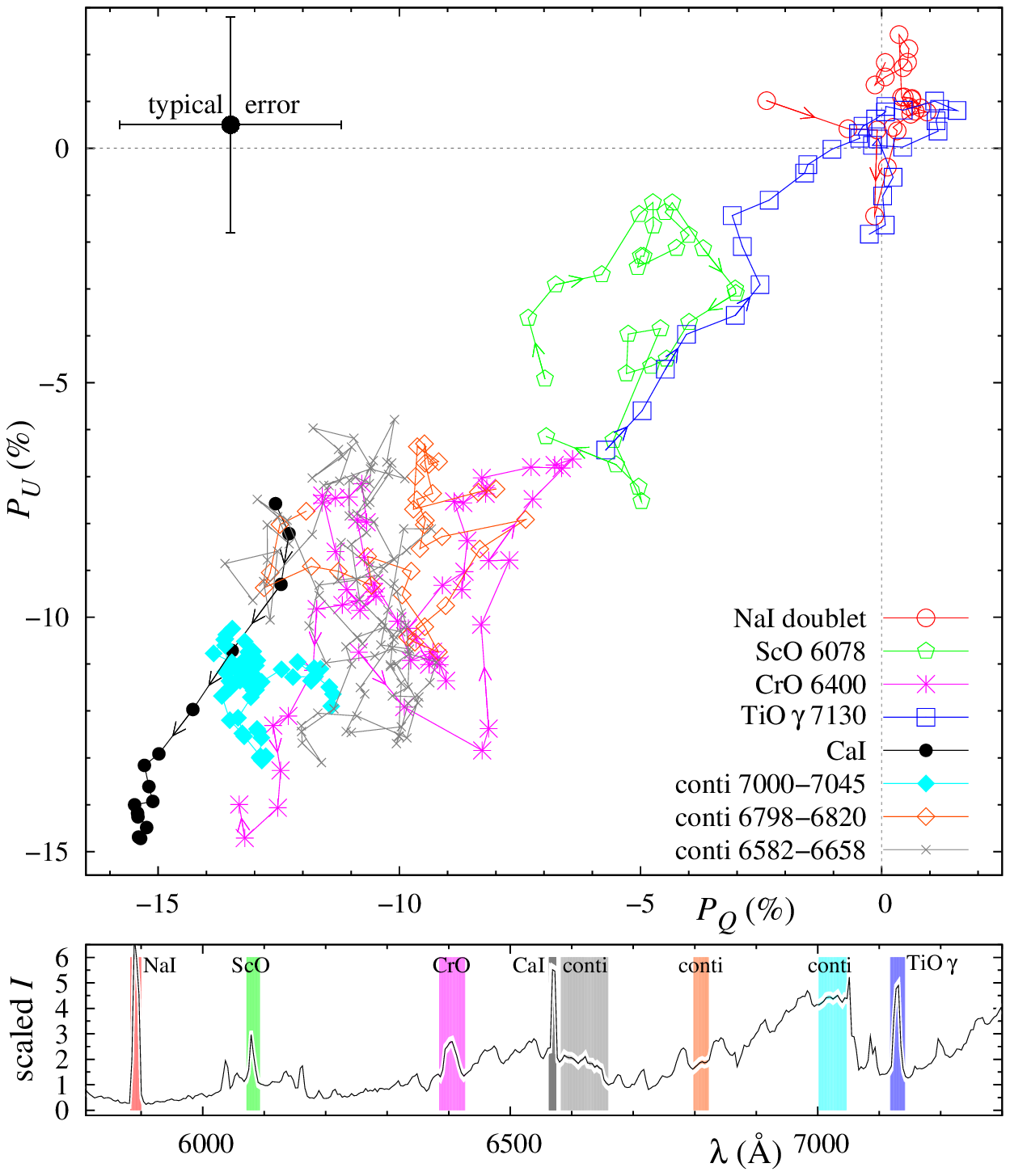}
\includegraphics[angle=0,width=0.5\textwidth]{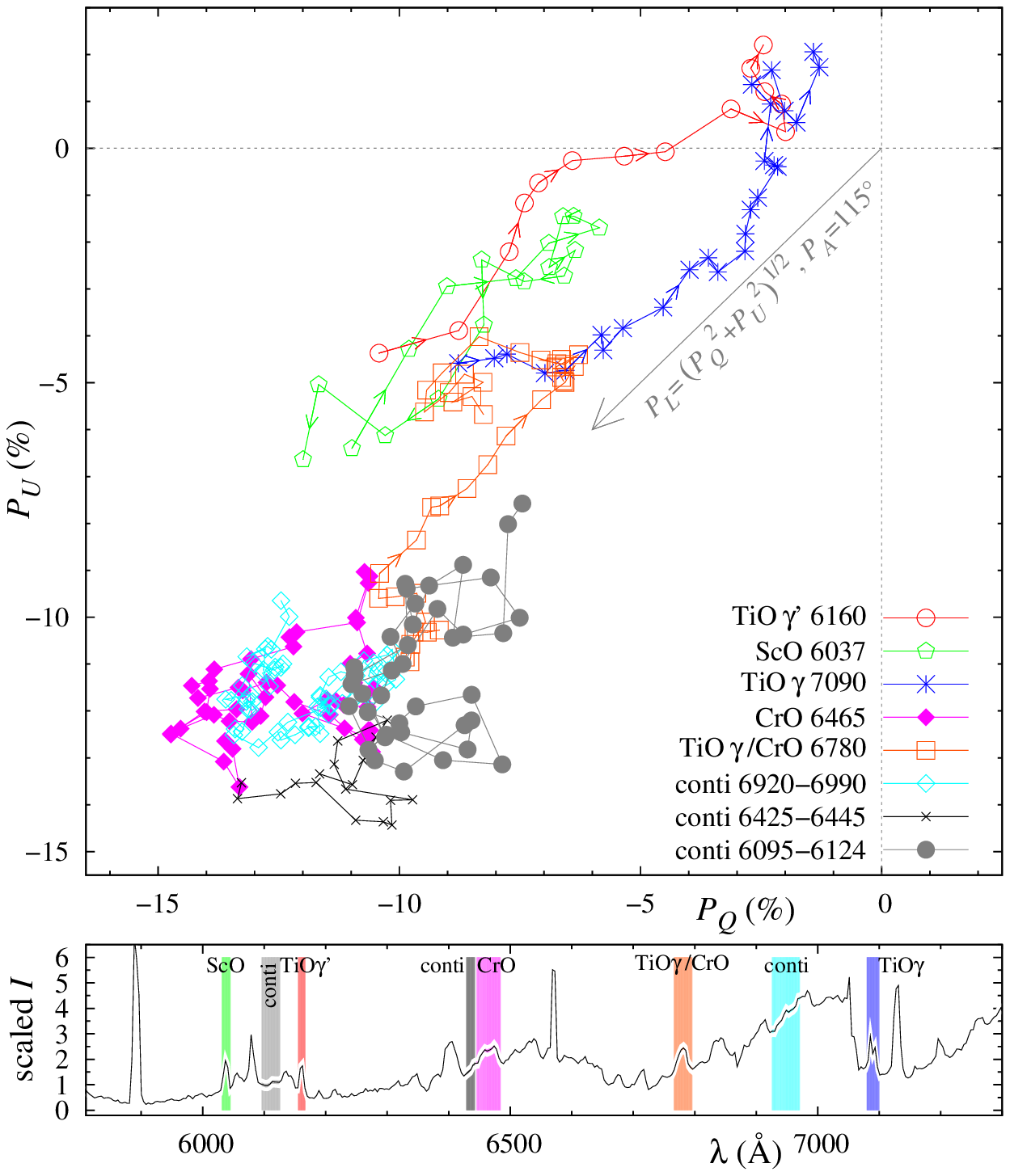}
\caption{$QU$ plots for most prominent emission features and arbitrary chosen continuum ranges in the spectrum of V4332 Sgr. The bottom panels show the spectral location of the features for which the Stokes parameters are displayed in the top panels. The arrows shown for some features show the direction of increasing wavelength. The dashed lines cross at the plots' origins marking the point of zero polarization.}
\label{fig-QU}
\end{figure*}

We also present results of our spectropolarimetric observations in terms of $QU$ plots in 
Fig.~\ref{fig-QU}. They present the two Stokes parameters, $P_Q$ and $P_U$, within the 
profiles of the most prominent emission features and the six other spectral ranges chosen 
to represent the continuum. These continuum samples were chosen in spectral ranges 
observed at high signal-to-noise ratios, outside telluric absorption bands, 
and are not considerably contaminated by other emission features (checked 
by comparing the total-power spectrum to the standard photospheric spectrum in 
Fig.~\ref{specpol2_fig}). The points representing emission features 
correspond to the flux of the emission feature \emph{and} the underlying 
polarized continuum. The Stokes parameters displayed in Fig.~\ref{fig-QU} are not 
corrected for the instrumental chromatic polarization (Sect.~\ref{obs_sect}). 
The points tracing the ``pure'' continuum occupy the high-polarization part 
of the $QU$ diagrams at around $(P_Q,P_L)\,=\,(-11,-13)$\%. The Stokes 
parameters of most of the emission features -- including the \NaI\ doublet, 
bands of TiO, and ScO -- show a depolarization pattern and are situated in 
the diagram between the continuum and the graph's origin. Additionally, 
within profiles of each of the
analysed bands of TiO, the polarization decreases with increasing wavelength, 
possibly owing to the changing contribution of the photospheric spectrum 
underlying the bands (cf. Fig.~\ref{specpol2_fig}; see Sect.~\ref{sect-inter}). 
The bands of CrO show essentially the same polarization as the continuum. 
Somewhat anomalous is the polarization pattern observed in the emission line 
of \CaI, whose blue wing has the Stokes parameters consistent with those of 
the continuum, but the net polarization increases with wavelength so steeply 
that the red wing shows polarization that is even higher than that of 
the photospheric spectrum. The changes in the Stokes parameters within 
the wavelength range covered by \CaI\ can again be ascribed to the slope 
in pseudo-continuum. In the next section, we interpret these polarization 
characteristics and try to separate the continuum polarization from 
depolarization effects in emission features.

\section{Analysis and interpretation}\label{sect-inter}
The observed strong polarization of the stellar-like continuum, and depolarization in
regions dominated by emission features is
consistent with the interpretation of the observed spectrum of
V4332~Sgr proposed in \citet{kst} and \citet{kt11}. According to the
hypothesis outlined in these references, the main object, resembling
an M-type giant, is surrounded by a dusty disc or disc-like envelope, most
probably formed during the 1994 eruption of V4332~Sgr. The
disc is seen almost edge-on, so the central object is not directly visible
for us. Some amount of the material lost during the eruption is flowing out more or
less along the disc axis, i.e. roughly perpendicularly to the line of sight.
The observed stellar-like continuum stems from scattering the light of 
the central object on dust grains in the disc and the outflow. The
scattering off the polar outflow, 
resulting at angles close to $90\degr$, is the main cause
of the observed polarization.
The atomic lines and molecular bands seen in emission result from
radiative excitation of atoms and molecules in the
circumstellar matter by the radiation of the central star. 
Therefore the emission features are not expected to show any significant 
polarization. 

Within the above scenario, the observed polarization, $P_{L\rm ,obs}$, 
in a given spectral region, where the observed flux is a sum of a polarized scattered
continuum and an unpolarized emission feature, can be calculated from
\begin{equation}
  P_{L\rm ,obs} = P_{L\rm ,sc} \frac{F_{\rm sc,em}}{F_{\rm em}+F_{\rm sc,em}},
\label{pol_eq}
\end{equation}
where $F_{\rm em}$ is the flux of the unpolarized emission feature while
$F_{\rm sc,em}$ is the flux of the scattered continuum underlying the
feature. The flux $F_{\rm sc,em}$ is polarized to a degree equal to $P_{L\rm ,sc}$. If 
$F_{\rm sc}$ denotes a mean flux of the observed continuum measured on both
sides of the emission feature, then Eq.~(\ref{pol_eq}) can be rewritten as
\begin{equation}
  P_{L\rm ,obs} = P_{L\rm ,sc} \frac{F_{\rm sc}}{F_{\rm em}+F_{\rm sc,em}}
      \frac{F_{\rm sc,em}}{F_{\rm sc}}.
\label{pol_eq2}
\end{equation}
Here, $F_{\rm sc}/(F_{\rm em}+F_{\rm sc,em})$ can be measured from 
the observed spectrum, although we cannot be sure that the flux measured 
next to the emission feature does not include weak emission features.
The $F_{\rm sc,em}/F_{\rm sc}$ ratio cannot be derived from the observations, 
but in most cases it can be assumed to be equal to 1.

The value of $P_{L\rm ,sc}$ is not easy derivable from the observed spectrum
because it is not straightforward to select spectral regions, 
which are absolutely free of emission features.
Local maximum values of the polarization degree (seen in 
Fig.~\ref{specpol_fig}), which can be regarded as corresponding to
the spectral places least contaminated by emission features, 
suggest a systematic decrease in the
polarization from $\sim$25\% at the short wavelengths to $\sim$17\%
at long wavelengths. These are, however, uncertain estimates, since the
signal-to-noise ratio of our observations is not very high and degrades when
going from long wavelengths to short ones. A decrease in the polarization 
degree with wavelength was also suggested in
\citet{kt11}, who obtained a higher polarization degree in the $V$
photometric band than in the $R$ band (26\% vs 11\%).

To obtain a more quantitative estimate of $P_{L\rm ,sc}$, we selected eight 
wavelength regions
where the spectrum seemed to be free of emission-line features. 
The mean polarization degree and angle derived from these regions 
are displayed in Fig.~\ref{specpol_fig}.
The error bars represent standard deviations of the
individual measurements from the mean values.  Error-weighted mean values of the
polarization degree and angle derived from all these selected regions
are $16.60\%\pm0.84$\% and $114\fdg6\pm2\fdg0$, respectively. 
If the selected regions
are affected by weak emission features,
then the polarization degree of the continuum would be higher than the
values we derived. 

\begin{figure}
\includegraphics[angle=270,width=\columnwidth]{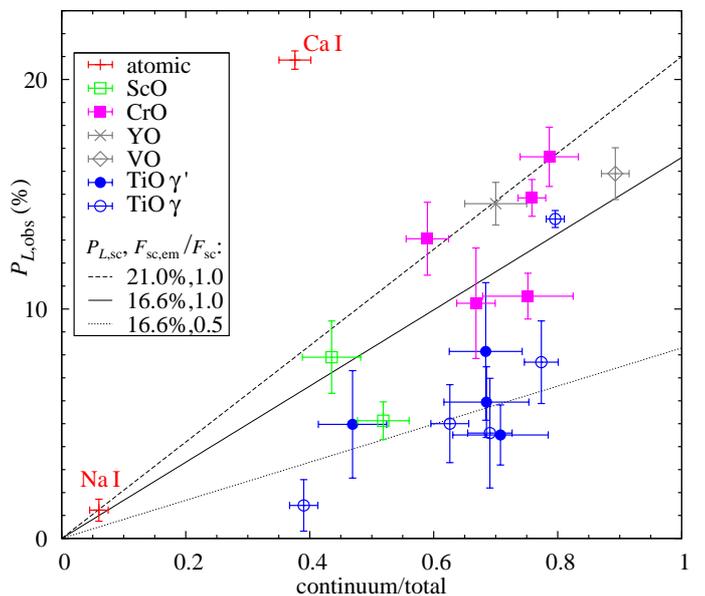}
\caption{Observed polarization degree of the emission features 
($P_{L\rm ,obs}$) as a function of the
ratio of the continuum flux to the total flux within the line profile
($F_{\rm sc}/(F_{\rm em}+F_{\rm sc,em})$). 
%Open circles:
%bands of TiO. Asterisks: bands of CrO. Leftmost symbol: \NaI\ doublet,
%uppermost symbol: \CaI\,$\lambda$6573 line. 
Lines show the relations calculated from Eq.~(\ref{pol_eq2}) for three
combinations of $P_{L\rm ,sc}$ and $F_{\rm sc,em}/F_{\rm sc}$. See text for
more details.}
\label{line_pol_fig}
\end{figure}

We measured mean fluxes and mean polarization degrees  
within the emission profiles, as well as the continuum level on both sides 
in the nearest vicinity of the features. The results are
shown in Fig.~\ref{line_pol_fig}, which presents the observed
polarization degree of a given emission feature
against the ratio of the adjacent continuum flux to the flux within
the feature, i.e., $F_{\rm sc}/(F_{\rm em}+F_{\rm sc,em})$ according
to the notation of Eq.~(\ref{pol_eq2}). The lines drawn in the figure show 
three relations calculated from Eq.~(\ref{pol_eq2}): 
the full line for  $P_{L\rm ,sc} = 16.6\%$ and $F_{\rm sc,em}/F_{\rm sc} = 1$;
the dashed one for $P_{L\rm ,sc} = 21.0\%$ and $F_{\rm sc,em}/F_{\rm sc} = 1$; and 
the dotted one for $P_{L\rm ,sc} = 16.6\%$ and $F_{\rm sc,em}/F_{\rm sc} = 0.5$.

Figure~\ref{line_pol_fig} demonstrates that the observed polarization 
characteristics of most emission features are well accounted for by the
hypothesis that the observed spectrum of V4332~Sgr stemss from
interactions of the radiation of the invisible central star with the
circumstellar matter. The observed stellar-like continuum results
from scattering on dust grains and is polarized at an average level of
$\sim$17\%. 

A somewhat higher polarization is observed in a few bands, mostly of CrO,   
which are situated above the full line in Fig.~\ref{line_pol_fig}.
These emission features may be partly produced near the central object. 
A fraction of their flux would then be scattered on dust, 
similarily as the photospheric spectrum, and thus would
contribute to the relatively high polarization of the observed features. The
polarized flux (see Fig.~\ref{specpol2_fig}),
which shows certain residuals of the CrO emission features,
supports this interpretation. 
This is also apparent in the $QU$ diagrams (Fig.~\ref{fig-QU}),
where CrO occupies the same parameter space as the continuum.

The case of the TiO bands, which mostly appear below the full line in 
Fig.~\ref{line_pol_fig}, can be explained by $F_{\rm sc,em}/F_{\rm sc} < 1.0$.
This means that the stellar continuum directly underlying the TiO
emission features is lower than what we measured as a continuum
next to the features. This is very likely given that we do
see deep and wide absorption bands of TiO in the stellar-like spectrum of
V4332~Sgr. The observed emission of TiO is formed near the heads of the
bands, where the stellar absorption features are expected to be deepest. Indeed, the
spectrum of polarized flux, as well as the standard M5\,III spectrum 
(Fig.~\ref{specpol2_fig}), show clear
``deepenings'' at the positions of the strongest TiO emission bands. 
As noted earlier, these steep changes in the continuum level can explain
the polarization pattern observed for TiO in the $QU$ graphs, 
namely that the net polarization decreases with wavelength 
at nearly constant polarization angle.

\subsection{The anomalous polarization of the \CaI\ line}
The general trend in Fig.~\ref{line_pol_fig} is that 
the stronger the emission line relative to
the continuum, the weaker its observed polarization. The only outlier from this
trend is the \CaI\,$\lambda$6573 line. 
This is the second strongest emission feature in our
spectrum (after the \NaI\,D lines), yet it shows the highest linear polarization
among the emission features, i.e. $20.8\% \pm 0.4\%$. This value is higher
than the mean polarization degree we obtained for 
the continuum (Fig.~\ref{specpol_fig}). It
is, however, very close to the continuum polarization used to plot 
the dashed line in Fig.~\ref{line_pol_fig} and close to the highest 
values that can be read from the middle panel of Fig.~\ref{specpol_fig}.
The polarization angle within the \CaI\ line is $113\fdg0\pm0\fdg5$, which
is practically the same value as for the stellar continuum 
(Fig.~\ref{specpol_fig}). 
Because the points in Fig.~\ref{specpol_fig} may 
underestimate the polarization degree 
of the continuum, we can conclude that the polarization pattern of  
the \CaI\,$\lambda$6573 line is, within the uncertainties of our observations,
the same as that of the stellar-like continuum of V4332~Sgr. 
This is also apparent in the $QU$ graph (Fig.~\ref{fig-QU}), 
where the Stokes parameters of the \CaI\  emission fall very close 
to the cluster of points corresponding to continuum, and only the 
red-shifted part of the \CaI\ profile drifts 
into the lower left-hand corner of the $QU$ plot.

A straightforward interpretation of such a polarization pattern could be that 
the \CaI\,$\lambda$6573 line, unlike the rest of the observed emission features,
is not produced by radiative excitation in the polar outflow 
but originates in the same way as the observed stellar continuum. Namely, it 
is emitted near the hidden central
object and then scattered on dust grains in the circumstellar matter. The
observed width of the \CaI\ line, which is about twice smaller than that of
the \NaI\ lines \citep{kst}, may suggest that emission of the two species 
does not arise from the same volume of gas (with emission of \NaI\ being more extended).
This interpretation, however, raises a question of why the \NaI~D lines are not
produced in the same way. Both the \NaI\ and \CaI\ lines require
similar excitation conditions (similar ionization and excitation energies in
both cases, similar abundances of Na and Ca), so if the \CaI\ line is 
emitted near the central object, the \NaI\ lines should there be produced as well 
--- in fact, even more easily than
the \CaI\ line, since the latter arises from an intercombination transition and
has a probability of spontaneous radiative decay that is four orders of magnitude lower
than the \NaI\ lines. Nevertheless, the \NaI\ lines show practically no polarization,
which implies that the contribution of the \CaI\ emitting region, hypothetically
placed in the immediate vicinity of the central object, to the
observed emission in the \NaI\ lines is negligible. 

That the \CaI\,$\lambda$6573 emission develops from an intercombination
transition prompts us to suggest that perhaps this is why the
line is so strongly polarized compared to the \NaI\ lines. 
The low probability for the radiative
de-excitation of the $4p\,^3P^o$ level (the upper level of \CaI\,$\lambda$6573)
favours conditions leading to an
overpopulation of the level if an effective excitation process exists. Then, it
would be easy to initiate a laser-like process leading to a fast
depopulation of the level through photon-stimulated transitions. If the
laser-like action is initiated by the stellar photons scattered on dust
grains, then all the photons emitted in the action would conserve the
polarization pattern of the scattered continuum, as
observed in the case of the \CaI\,$\lambda$6573 line. The problem, however, 
is that we cannot propose any physical process overpopulating 
the $4p\,^3P^o$ level of \CaI. In
addition, regions where \CaI\ would mainly be at this level should be hidden
from the direct stellar radiation. Otherwise the laser-like action would be
initiated by these direct photons rather than by those scattered on dust.

\section{Conclusions}
The optical spectrum of V4332~Sgr shows a significant polarization. It is
primarily the stellar-like continuum that is strongly polarized. We
interpret this as evidence that we do not directly observe the
central object but rather the effect of scattering of its spectrum on dust
grains in the circumstellar matter. Most of the molecular bands in emission and the
\NaI\ D emission lines are not polarized, since they are produced in the
circumstellar gas by radiative excitation due to absorption of photons
from the central star. These emission features are therefore observed directly
from the polar outflow. The bands of CrO and the line of \CaI\ show intrinsic 
polarization characteristics that are similar to that of the photospheric
spectrum, and therefore  their emission is mostly produced close to the photosphere. 
Their radiation is not seen directly, because it is mostly hidden by the disc  
and reaches the observer only owing to scattering in the disc and in the polar outflow.

The spectrum of V4332~Sgr is the only instance where emission bands of CrO have 
been observed \citep{kst}. This suggests some abundance peculiarities in 
the molecular envelope of this exotic object. Because the emission bands 
are excited by radiation, the excitation temperatures derived from 
optical observations \citep{kst} do not constrain the location of 
the CrO gas in the complex system of V4332~Sgr. The spectropolarimetric 
observations presented here locate this gas close to the star, thus providing 
rough but significant constraints for any model attempting to explain 
the chemical peculiarity of this object.

Prior to this work, only broad-band polarimetric measurements of V4332~Sgr 
were known. Although they revealed the high degree of linear polarization 
of the optical light, no information about the potential depolarization 
effects in emission lines was available, hampering realistic attempts 
to fully characterize the geometries of the scattering medium and the 
emitting gas. Here we were able to identify the polarization patterns 
in both lines and continuum, showing that the spectropolarimetric 
observations bear important constraints on the system's geometry. 
This opens a way to construct a more detailed  and quantitative model 
of the circumstellar environment of V4332~Sgr. Such modelling effort is 
pending. Knowing both the physical and chemical structure of the 
circumstellar vicinity of  V4332~Sgr is important for better understanding 
of stellar-merger events and their aftermath. 

Although polarization studies of emission features in dusty circumstellar
environments of late-type stars are very sparse, 
(de-)polarization effects similar to those
described here have been reported in the carbon stars
R\,CrB and V854\,Cen, where the \ion{Na}{I} doublet and molecular bands of C$_2$, both seen
in emission, show a significant drop in polarization degree 
with respect to a polarized continuum \citep{bieging,whitney}.
The polarization effects are much more pronounced in V4332~Sgr,
making it a very unique object for future spectropolarimetric studies of dusty envelopes.

\begin{acknowledgements}
We thank our referee, D. Harrington, for constructive comments that helped 
us improve the data presentation in this paper.
The research reported here has partly been supported from a grant no.
N\,N203\,403939 financed by the Polish National Science Centre.
\end{acknowledgements}

%--------------------------------------------------------------------------------
\begin{appendix}
\section{Interstellar polarization}\label{ISP}
The FORS2 slit, which consists of 15 ``slitlets'', was aligned to a zero 
position angle. By chance, this position  allowed registering a few 
field stars simultaneously with V4332~Sgr (although not all of 
the stars were perfectly aligned on the slit). Polarization parameters 
for the brightest of them are presented in Table\,\ref{fieldstars}. 
The polarization angle could only be determined reliably for one star. 
The polarization degree of the listed stars is fairly high, possibly 
indicating that the interstellar polarization in the field is relatively strong. 
The limited number of measured stars and unknown distances do not allow 
for strong constraints on the interstellar polarization towards V4332~Sgr, 
although it is likely to be $\lesssim$3\%. With the relations of 
\citet{serkowski} and the interstellar reddening of the object of 
$E_{B-V}$=0.32\,mag \citep{kst}, we obtain the same upper limit on 
the interstellar polarization of V4332~Sgr, i.e., $P_L^{\rm ISM}\lesssim$3\%. 
Within the uncertainties, these limits are consistent with 
the lowest polarization degree observed in V4332~Sgr 
(within the profile of the \ion{Na}{I} doublet). 
%As discussed in the main text,  the polarization of V4332~Sgr is considerably larger than this interstellar component and therefore it can be safely neglected in the analysis.

%
\begin{table}
\caption{Linear polarization of measured field stars.}\label{fieldstars}
\begin{tabular}{cccc}
\hline
Declination\tablefootmark{a}&
Distance\tablefootmark{b}& 
$P_L$ (1$\sigma$) & $P_A$ (1$\sigma$)\\
&(arcmin)&(\%)&\\
\hline
--21:23:38.3&~0.17&1.81 (0.98)&\\
--21:22:50.4&~0.68&0.68 (0.76)&71\fdg2 (1\fdg0)\\
--21:21:05.9&~1.80&3.00 (0.58)&\\
--21:25:37.8&--2.15&1.44 (0.87)&\\
\hline   
\end{tabular}  
\tablefoot{
\tablefoottext{a}{Right ascension is approximately the same (within $\sim$0\fs1) as that of V4332~Sgr, $\alpha_{2000}$=18:50:36.70.}
\tablefoottext{b}{Angular offset from V4332~Sgr in declination, positive for north, negative for south.}}
\end{table}
%(Its PA is baised by the instrumrntal PA, which changes over the FORS2 field of view)
%-------------------------------------------------------
\end{appendix}
%-----------------------------------------------------------------------------------------------------  
  
%\bibliographystyle{aa}

\end{document}